\documentclass[aps,prl,reprint,groupedaddress]{revtex4-1} 
 
\usepackage{graphicx} 
\begin{document} 
 
\title{Nuclear correlations and the r-process} 
 
\author{A. Arcones} 
 
\email[]{a.arcones@unibas.ch} 
 
\affiliation{Department of Physics, University of Basel, 
  Klingelbergstra{\ss}e 82, CH-4056, Switzerland}

\author{G. F. Bertsch} 

\affiliation{Department of Physics and Institute for Nuclear Theory,
  Box 351560, University of Washington, Seattle, Washington 98915,
  USA}
 
\date{\today} 
 
\begin{abstract} 
  We show that long-range correlations for nuclear masses have a
  significant effect on the synthesis of heavy elements by the
  r-process. As calculated by Delaroche et al.
  \cite{Delaroche.etal:2010}, these correlations suppress magic number
  effects associated with minor shells.  This impacts the calculated
  abundances before the third r-process peak (at mass number $A\approx
  195$), where the abundances are low and form a trough. This trough
  and the position of the third abundance peak are strongly affected
  by the masses of nuclei in the transition region between deformed
  and spherical. Based on different astrophysical environments, our
  results demonstrate that a microscopic theory of nuclear masses
  including correlations naturally smoothens the separation energies,
  thus reducing the trough and improving the agreement with observed
  solar system abundances.
\end{abstract}

 
\maketitle

The rapid neutron capture process (r-process) leads to half of the
heavy elements beyond iron. In this nucleosynthesis process many
neutrons are captured by seed nuclei (i.e., iron group and nuclei up
to $A\approx 90$) in time scales shorter than beta decay. The pathway
for the r-process thus involves extremely neutron-rich nuclei far from
stability and close to the neutron drip-line. However, only few of
these very exotic nuclei can be produced in current rare isotope beam
facilities.  Even with next-generation facilities not all nuclei
relevant for the r-process will be reached.  In view of the
experimental difficulties, calculations based on theoretical models of
the reaction chain are essential to understand nucleosynthesis and the
origin of the heaviest nuclei in the universe.

Even more than the nuclear physics uncertainties, the astrophysical
environment of the r-process is far from understood
\cite{arnould.goriely.takahashi:2007}. The rapidity and the high
neutron densities required for this process point to the most
neutron-rich objects in the universe: neutron stars. Many studies have
been performed for the high-entropy, neutrino-driven wind subsequent
to core-collapse supernovae (see \cite{arnould.goriely.takahashi:2007}
for a review). The conditions necessary for the r-process have been
well established but are not found in recent hydrodynamic simulations
\cite{arcones.janka.scheck:2007, Huedepohl.etal:2010,
  Fischer.etal:2010, Arcones.Janka:2011}. The merger of two neutron
stars was proposed in \cite{Lattimer.Schramm:1974} and indeed
simulations can reach the extreme conditions required for the
r-process (see, e.g., \cite{Freiburghaus.etal:1999,
  Goriely.etal:2011}). However, neutron star mergers are not occurring
early enough in our galaxy to explain the heavy r-process elements in
very old stars \cite{Argast04}.

Typically, the r-process is modeled taking as theoretical input some
scenario of the astrophysical environment together with specific
models for nuclei where experimental information is not available.
While individual models can be judged by the quality of agreement for
the predicted final abundances, the dependence on the nuclear physics
entering the different models and which physics features cause
discrepancies is not fully understood.  In this Letter, we show
that some features of the final abundances can be traced to the
treatment of correlations in the calculated masses.  It has been found
that long-range correlations associated with quadrupole degrees of
freedom can significantly improve the description of nuclear masses
with respect to Hartree-Fock-Bogoliubov (HFB) theory, using either a
Skyrme interaction~\cite{ben06} or the Gogny
interaction~\cite{Delaroche.etal:2010}.  It was also found that the
correlation energy enhances magic number effects at double shell
closures and weakening them elsewhere \cite{ben08}, a phenomenon known
as ``mutually enhanced magicity'' \cite{lu03}.  The mass table of
Ref. \cite{Delaroche.etal:2010} contains both the HFB and the
correlation contribution permitting us to examine its specific effect
on nucleosynthesis.  In both Refs.~\cite{Delaroche.etal:2010} and
\cite{ben06} the HFB theory was extended by the generator coordinate
method to treat quadrupole correlations, and no parameters were
assumed beyond those going into Skyrme or Gogny energy functionals.
 
Since there is uncertainty about the astrophysical environment of the
r-process, we study the effects of long-range nuclear correlations for
the different scenarios mentioned above.  Specifically, we consider
trajectories arising from hot or cold, high-entropy neutrino-driven
winds \cite{arcones.janka.scheck:2007} and from neutron star mergers
\cite{Rosswog99}. In all cases we find that correlations reduce the
trough in the abundances before the third r-process peak (at
$A\approx195$). Similar reductions have been achieved in
nucleosynthesis calculations based on mass models with shell quenching
\cite{Chen.etal:1995}. Here we show that this trough depends not only
on the shell closures, but also on the transition from deformed to
spherical nuclei.

\begin{figure*}[!hbt] 
  \includegraphics[width=0.5\linewidth]{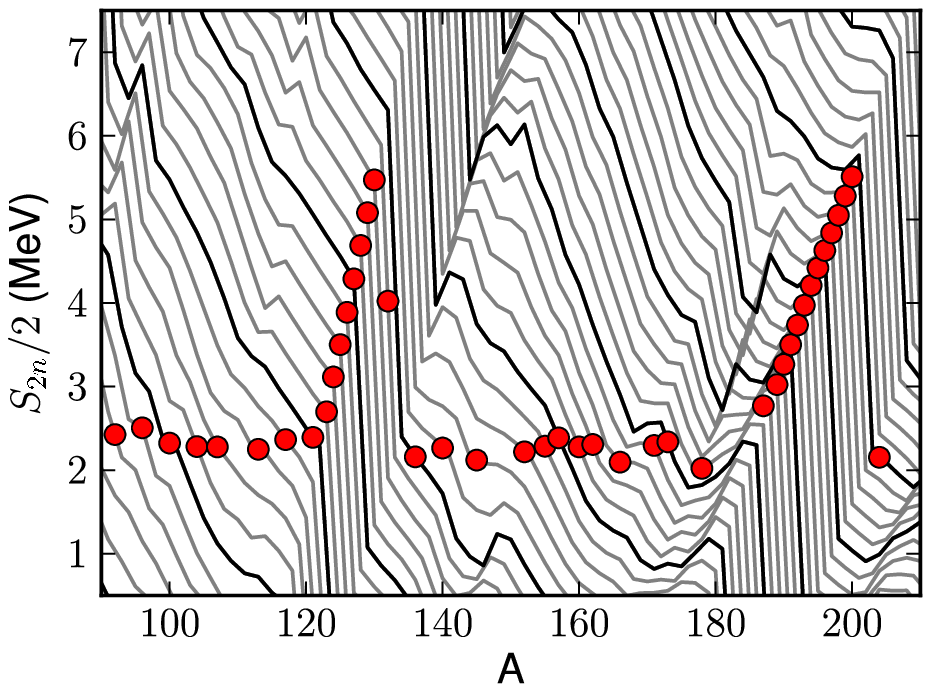}%
  \includegraphics[width=0.5\linewidth]{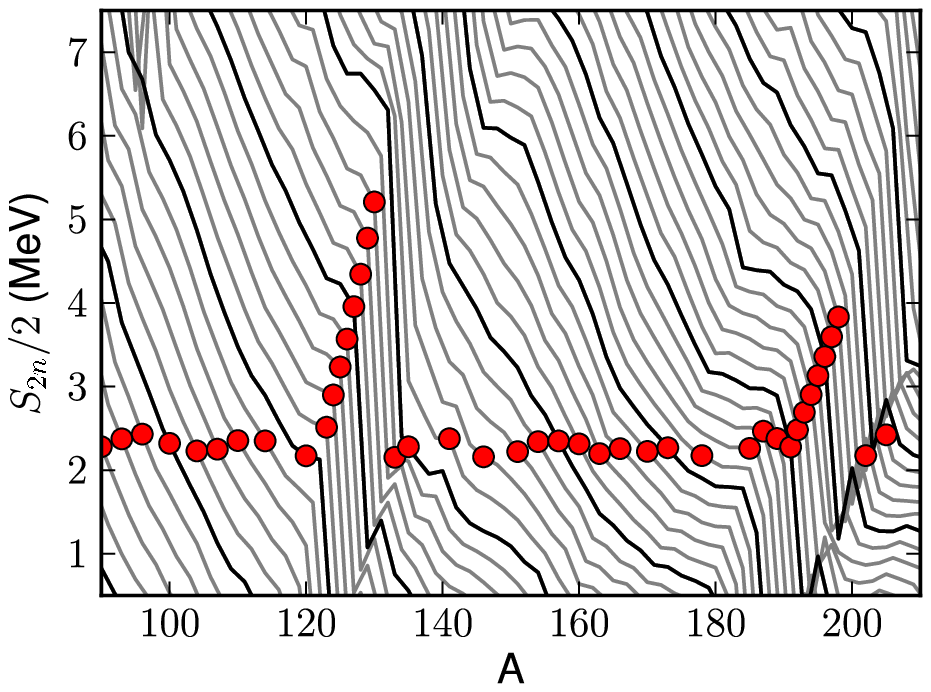} 
  \caption{(Color online) Half the two-neutron separation energy
    $S_{2n}/2$ for constant proton number as a function of mass number
    $A$. The lines represent isotopic chains from $Z=30$ to
    $Z=80$. The dots mark the r-process path for the hot wind
    trajectory at freeze-out. The left panel corresponds to the
    nuclear masses of Delaroche et al.~\cite{Delaroche.etal:2010}
    without nuclear correlations. In the right panel nuclear
    correlations are included.}
  \label{fig:s2n} 
\end{figure*}

The most important input to the reaction network from the perspective
here is the neutron separation energy, which crucially affects the
$(\gamma,n)$ photodissociation reactions and the $(n,\gamma)$ capture
rates.  The separation 
energies obtained from the mass tables of Delaroche, el al.  
\cite{Delaroche.etal:2010} are shown in Fig.~\ref{fig:s2n}.  The
lefthand and righthand panels correspond respectively to calculated masses
in HFB theory and in their extended theory including correlations.  Plotted are 
contours at constant proton number of half the two-neutron separation
energy $S_{2n}$, as a function of mass number.  
Delaroche et al.~\cite{Delaroche.etal:2010} calculated only
even-even nuclei, so the needed odd-A separation energies will be
assigned by interpolation.  Also, their correlated masses are higher
in some nuclei and for those masses we use the HFB values.
With the theoretical $S_n$ values, we
calculate the neutron-capture cross sections with a simplified
Hauser-Feshbach model~\cite[Eq.~(39) in
Ref.][]{Woosley.Fowler.ea:1975}.  The nuclear binding energy
differences also affect beta-decay rates, but the sensitivity is not
as high \cite{Arcones.MartinezPinedo:2011} and we use the table of
\cite{Moeller.Pfeiffer.Kratz:2003} for these transitions.

We point out two features in the calculated two-neutron separation
energies (Fig.~\ref{fig:s2n}) that have a major effect on the
r-process abundances, see also~\cite{Arcones.MartinezPinedo:2011} and
references therein.  First, the abrupt drop of $S_{2n}$ around $A=130$
and $A=195$ corresponds to the magic numbers $N=82$ and $N=126$,
respectively. For these nuclei with closed shells, neutron-capture
cross sections are very small and the photodissociation rate is high
for typical r-process temperatures. As a result the r-process path
stops at these nuclei and waits for beta decay. It is here that matter
accumulates and the abundance peaks form, as they are observed in the
solar system. The second characteristic feature is the smoothness of
the topography just before the magic numbers.  Without correlations,
the evolution of $S_{2n}$ with $A$ shows a pronounced dip just before
the $N=126$ magic number, associated with the transition from deformed
to spherical nuclei. This behavior is also present in other mass
models \cite{Moeller.Nix.ea:1995, Pearson.Nayak.Goriely:1996,
  Goriely.Chamel.Pearson:2009}.
 
Nuclear correlations strongly affect these two features. When
correlations are included, the neutron shell closures are smoother,
similarly to what occurs when introducing shell
quenching~\cite{Moeller.Pfeiffer.Kratz:2003}. The less pronounced drop
of $S_{2n}$ at $N=82$ and $N=126$ leads to smaller peaks in
abundances.  The other important effect is the smoothing of the dip in
$S_{2n}$ just before $N=126$ ($A\approx 180$). This has a big impact
on late neutron captures and on the formation and evolution of the
trough in the abundances before the $A=195$ peak.
 
\begin{figure}[!htb] 
  \includegraphics[width=0.9\linewidth]{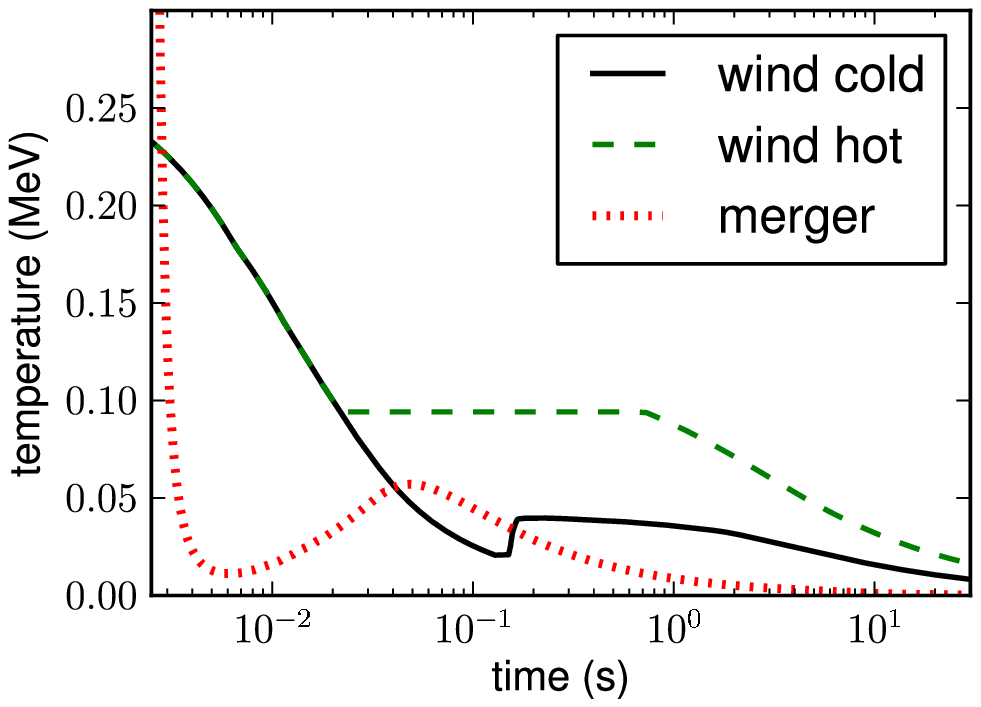}\\ 
  \includegraphics[width=0.9\linewidth]{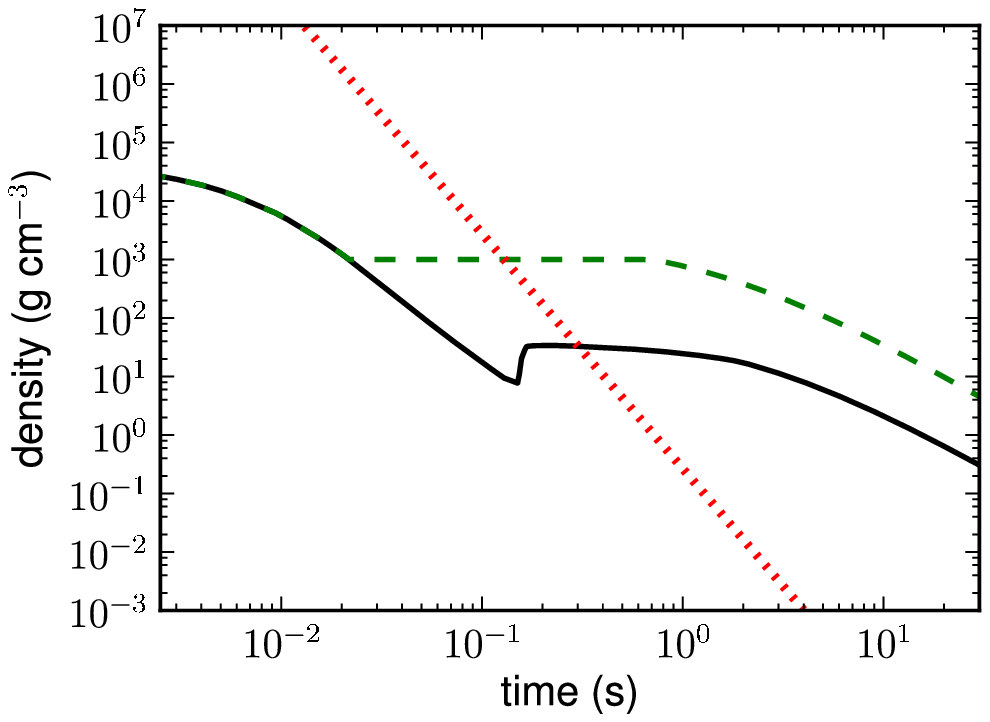} 
  \caption{(Color online) Evolution of temperature and density for
    different astrophysical environments. The trajectories for a
    neutrino-driven wind with cold (solid black line) and hot (dashed
    green line) r-process are based on the simulations
    of~\cite{arcones.janka.scheck:2007}. The trajectory for the
    neutron star merger~\cite{Freiburghaus.etal:1999} is given by
    dotted red line. \label{fig:traj}}
\end{figure}

The evolution of the astrophysical environment is shown in
Fig.~\ref{fig:traj} for the three scenarios we consider.  First, we
take two trajectories of neutrino-driven wind simulations
\cite{arcones.janka.scheck:2007} with the entropy artificially
increased to produce the third r-process peak (i.e., the same
trajectories as in \cite{Arcones.MartinezPinedo:2011}). The solid
black line represents a cold wind evolution where photo-dissociation
is negligible and the evolution is characterized by a competition
between neutron captures and beta decays.  The other trajectory,
assuming higher temperatures and densities (dashed green line), is
closer to a classical r-process \cite{Kratz.Bitouzet.ea:1993} in which
$(n,\gamma)-(\gamma,n)$ equilibrium is achieved. Moreover, we consider
a neutron star merger trajectory (dotted red line) from the
simulations of Ref. \cite{Rosswog99} as applied to the r-process (see
\cite{Freiburghaus.etal:1999}). Notice the fast drop of the density
(initially very high) and the small increase of the temperature (at
$t\approx10^{-2} - 1$~s) due to the energy released by r-process
reactions \cite{Metzger.Arcones.etal:2010}.
 
\begin{figure}[t]
  \includegraphics[width=0.9\linewidth]{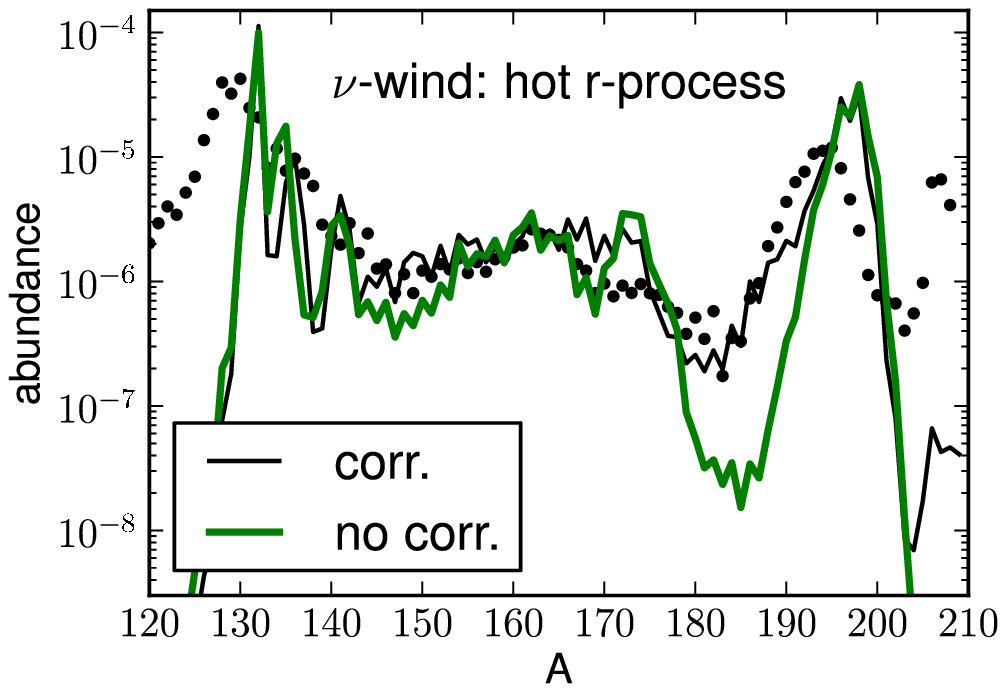}\\ 
  \includegraphics[width=0.9\linewidth]{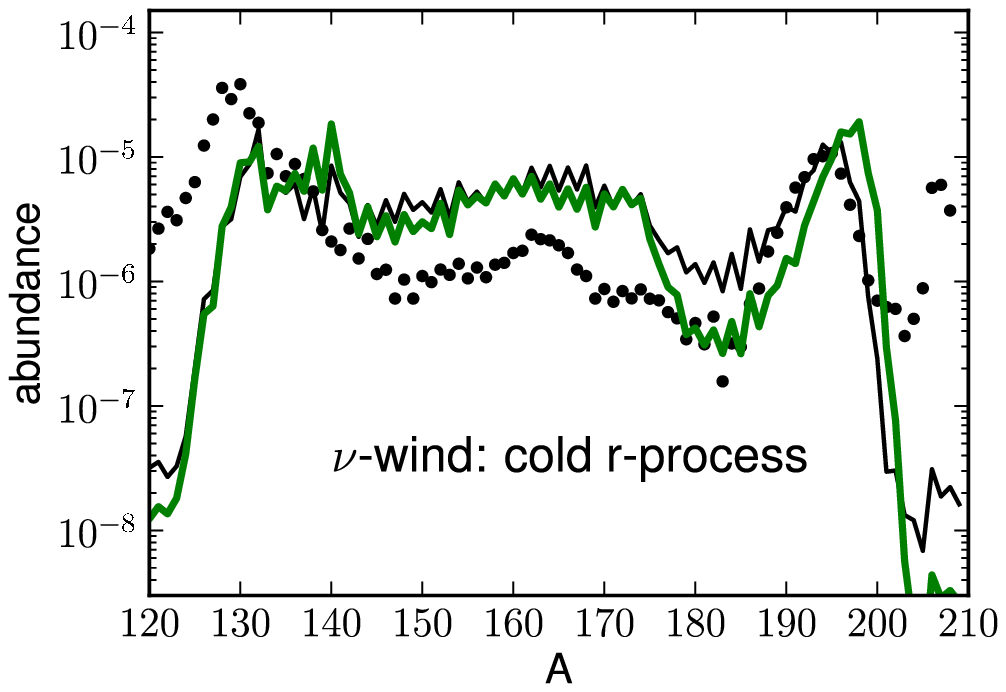}\\ 
  \includegraphics[width=0.9\linewidth]{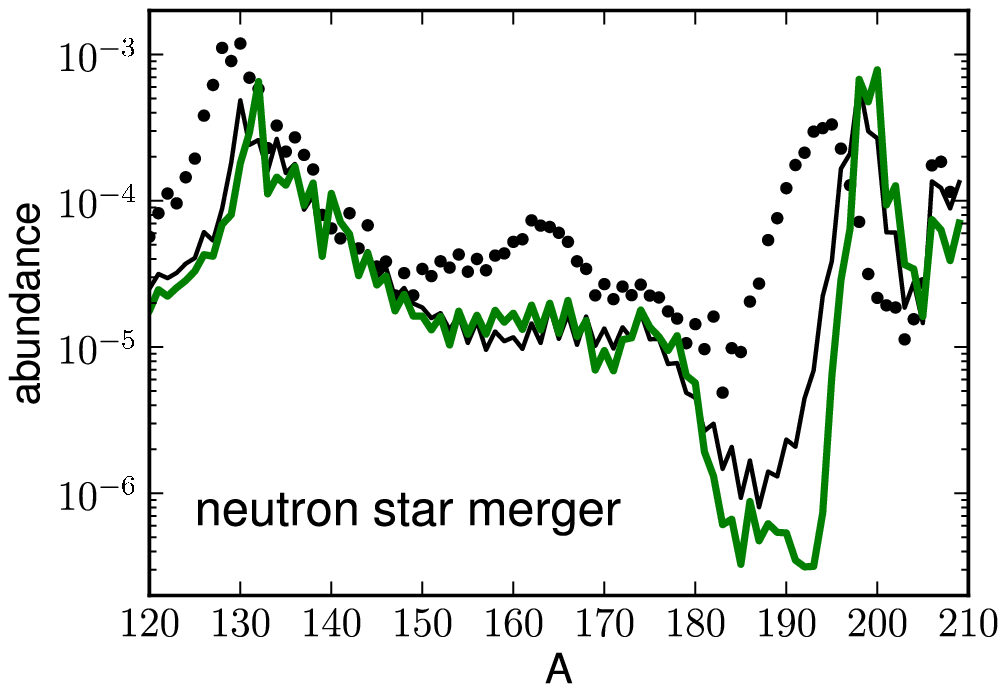} 
  \caption{(Color online) Abundances with and without nuclear
    correlations based on hot and cold wind and neutron star merger
    trajectories, from the top to the bottom. The black dots are solar
    system r-process abundances. \label{fig:abund}}
\end{figure}

The resulting abundances are shown in Fig.~\ref{fig:abund}.  The
nucleosynthesis calculations start at high temperature ($\sim 1$~MeV)
with nuclear statistical equilibrium. The initial evolution until
charged-particle freeze-out is calculated with a full network of
reactions and including nuclei up to
$Z=63$~\cite{Arcones.MartinezPinedo:2011,
  Freiburghaus.Rembges.ea:1999}. The r-process phase which follows
(after $T\sim 0.25$~MeV) is modeled with a reduced
network~\cite{Arcones.MartinezPinedo:2011,
  Freiburghaus.Rembges.ea:1999} that includes the relevant reactions
for the r-process (neutron capture, photodissociation, beta decay,
alpha decay, and fission). The masses from
Ref.~\cite{Delaroche.etal:2010} are used in this r-process network.
 
The abundances are very different for the three trajectories (see
Fig.~\ref{fig:abund}) because they depend very much on the temperature
and density evolutions.  For example, in the hot wind scenario
$(n,\gamma)-(\gamma,n)$ equilibrium is important in determining the
abundances.  In the other scenarios, the main competition is between
neutron captures and beta decays. The red dots in Fig.~\ref{fig:s2n}
show the r-process path at freeze-out (neutron-to-seed ratio equal
one) for the hot wind trajectory. The abundances at this point thus
provide crucial information for the r-process phase.  Afterwards
nuclei beta decay to stability, only changing the mass number by
beta-delayed neutron emission or by capture of the last neutron (see,
e.g.,~\cite{Arcones.MartinezPinedo:2011, Mumpower.etal:2011}). These
late-stage reactions lead to the stagger in the resulting abundances
between peaks. The observed abundances are much smoother; the
difference may be attributed to the approximation used for the neutron
captures~\cite{Arcones.MartinezPinedo:2011}.  None of the scenarios
shown here fit the entire range of solar system abundances.  Very
likely, the observed abundances arise from a superposition of
trajectories that individually emphasize the lighter or the heavier
mass numbers.  Our goal here was simply to show relative changes due
to correlations, in scenarios that lead to the heavier elements.
 
From Fig.~\ref{fig:abund} one sees that nuclear correlations affect
mainly the abundances in the region before the third peak
($180<A<190$). Here, the three trajectories show a similar behavior;
the correlations reduce the trough. However, the depth of the trough
and the impact of correlations depends on the trajectory.  For the hot
wind (top panel), $(n,\gamma)-(\gamma,n)$ equilibrium is reached and
the r-process path is given by the neutron separation
energy. Therefore, the features in $S_{2n}$ become crucial to
understand the final abundances. The abundances at freeze-out already
show a pronounced trough that is reduced and shifted toward higher $A$
during the decay to stability.  The evolution and fill-up of the
trough therefore depends on neutron captures. These can be enhanced
due to the features shown in $S_{2n}/2 \approx 3-4$~MeV at $A\approx
180$ without correlations (left panel Fig.~\ref{fig:s2n}). For the
cold wind (middle panel) the trough and the second peak are smaller
than for the hot wind. Because of the lower temperatures during the
r-process in this case the photodissociation is almost
negligible. Therefore, the impact of the masses is smaller as it
enters mainly through the neutron captures. The merger trajectory
(bottom panel) presents extreme abundances with a very narrow third
peak. The dip in $S_{2n}$ increases the neutron captures in the region
before the third peak and moves matter towards higher mass
numbers. This results in a narrower trough for the calculation with
correlations.

In summary, correlations smooth out the nucleus-to-nucleus variations
of nuclear separation energies.  This is especially pronounced for the
region before $N=126$ as shown by the two neutron separation
energy. The dip observed in $S_{2n}$ without correlations is also
present in other mass models~\cite{Moeller.Nix.ea:1995,
  Pearson.Nayak.Goriely:1996, Goriely.Chamel.Pearson:2009} (see Fig.~6
of~\cite{Arcones.MartinezPinedo:2011}).  In the classical r-process
\cite{Kratz.Bitouzet.ea:1993}, the problem of the large trough before
peaks was previously overcome by quenching the shell
gap~\cite{Chen.etal:1995}.  The masses of nuclei in the transition
region ($180<A<190$) between deformed and spherical strongly affect
the calculated abundances determining the shape and position of the
third peak and the trough before this.  We have shown that a
microscopic theory of nuclear masses including correlations is
sufficient to smoothen the separation energies and thus reduce the
trough in the abundances. Our study here shows the strong impact of
correlations on r-process nucleosynthesis and motivates further
theoretical work to treat the nuclear physics as realistically as
possible. For the future, we plan to study nucleosynthesis using
globally calculated beta decays and neutron captures consistently with
the theoretical separation energies.
  
\vspace{0.3cm}

\begin{acknowledgments} 
  We thank F.-K.~Thielemann, G.~Mart\'inez-Pinedo, and
  T.R.~Rodr\'iguez for discussions and comments on the
  manuscript. A.A. is supported by a Feodor Lynen Fellowship of the
  Alexander von Humboldt Foundation and by the Swiss National Science
  Foundation. G.F.B.  acknowledges support from the US Department of
  Energy under grant DE-FG02-00ER41132. We also thank the Institute
  for Nuclear Theory, where this work was initiated.
\end{acknowledgments} 
 
%
 
\end{document}